\documentclass[aps,pra,longbibliography,superscriptaddress,floatfix,twocolumn]{revtex4-1}

\usepackage{graphicx}
\usepackage{latexsym}
\usepackage{graphicx}
\usepackage{times}
\usepackage{color}
\usepackage{amsmath}
\usepackage{dcolumn}
\usepackage{mathbbol}
\usepackage{latexsym,amsmath,amssymb,bm,euscript}
\usepackage{xr}
\usepackage{lipsum}
\usepackage[normalem]{ulem}
\usepackage{relsize}
\usepackage{hyperref}
\hypersetup{
    colorlinks,%
    citecolor=blue,%
    filecolor=blue,%
    linkcolor=blue,%
    urlcolor=blue
}

\newcommand{\tr}{\text{Tr}}

\begin{document}
\title{Quantum Otto Thermal Machines Powered by Kerr Nonlinearity}

\author{Udson C. Mendes}
\address{Instituto de F\'{i}sica, Universidade Federal de Goi\'{a}s, 74.001-970, Goi\^{a}nia - Go, Brazil}
\author{Jos\'{e} S. Sales}
\address{Campus Central, Universidade Estadual de Goi\'{a}s, 75132-903, An\'{a}polis - Go, Brazil}
\author{Norton G. de Almeida}
\address{Instituto de F\'{i}sica, Universidade Federal de Goi\'{a}s, 74.001-970, Goi\^{a}nia - Go, Brazil}

\date{\today}

\begin{abstract}
We study the effect of Kerr nonlinearity in quantum thermal machines having a Kerr-nonlinear oscillator as working substance and operating under the ideal quantum Otto cycle. We first investigate the efficiency of a Kerr-nonlinear heat engine and show that by varying the Kerr-nonlinear strength the efficiency surpasses in up to 2.5 times the efficiency of a quantum harmonic oscillator Otto engine. Moreover, the Kerr-nonlinearity makes the coefficient of performance of the Kerr-nonlinear refrigerator to be as large as 3 times the performance of quantum harmonic oscillator Otto refrigerators. These results were obtained using realistic parameters from circuit quantum electrodynamics devices formed by superconducting circuits and operating in the microwave regime.
\end{abstract}

\maketitle

\section{Introduction}

Quantum thermal machines (QTM) produce or consume energy by using quantum matter as working substance~\cite{Dubois59,key-7Kosloff00,key-8Niewen05,key-5Mahler07,Nori07}. Depending on the nature of the quantum matter, bosons or fermions, and the reservoir that they are coupled, QTMs can, under certain conditions, surpass the efficiency of a classical Otto engine and even the efficiency of a classical Carnot engine. For instance, thermal engines and refrigerators in which conventional heat reservoirs are replaced by non-classical reservoirs~\cite{RefrigLong15,Manzano16,key-14Taysa19}, when operating either in the quasi-static limit and in the non-zero power limit, can surpass their classical analogues, showing efficiency larger than those obtained in the Carnot cycle~\cite{surpLutz14,key-11Klaers17}. Recently, it was shown that it is possible to surpass Carnot efficiency, in the case of engines, and Carnot performance, in the case of refrigerators, even far from the quasi-static regime~\cite{squeezXiao2018,Wang19,Sales20,key-15Rogerio20}.

In absence of quantum resources, in the form of either coherence or correlations between the quantum matter and the reservoirs with which the work substance interacts \cite{key-28Reeb14,key-27Bera19}, the formalism developed by Alicki~\cite{key-26Alicki79} provides an unambiguous definition for the quantum equivalents of heat and work. In this formalism, heat is defined as the amount of energy exchanged between the working substance and the thermal reservoir. In turn, work is defined as the change in the Hamiltonian during the cycle. This formalism was used, for example, to study an Otto engine interacting with a reservoir with effective negative temperature, which revealed the intriguing phenomenon of increased performance of an Otto quantum engine concomitant with the increasing power~\cite{Assis19}. In addition to the use of non-classical reservoirs, techniques such as shortcut to adiabaticity have been exploited to increase the performance of thermal machines \cite{abah-pre2018,deng-sadv2018,abah-prr2020}. In this case, QTMs are engineered to obtain the same performance of the quasi-static cycle using machines operating at non-null power \cite{Adiab_Muga19}.

Otto machines based on quantum harmonic oscillator (QHO) interacting with non-classical reservoirs and operating at finite time have been extensively investigated in the recent years. However, while implemented in real devices, quantum non-harmonic effects may not only be relevant, but also lead to new phenomena. In the context of quantum optics, nonlinear interactions are responsible, for example, to generation of non-classical states of light such as squeezed states. An important nonlinear effect is the Kerr non-linearity, which appears naturally in Josephson junction-based devices, among others. Indeed, Kerr nonlinearity is at the heart of quantum information devices based on superconducting circuits \cite{blais-rmp2021}. For instance, Kerr nonliterary is essential for the operation of the transmon qubit \cite{koch-pra2007} and it is used to stabilize cat-states qubits \cite{grimm-nature2020}. 

In this research paper, we investigate the efficiency $\eta$ and coefficient of performance $\epsilon$ of a cyclic QTM formed by a Kerr-nonlinear oscillator (KNO) by operating an Otto cycle. We show that the efficiency and performance of a Kerr-nonlinear Otto machine outperforms the efficiency and performance of its linear QHO counterpart.

The paper is organized as follow. In Sec.~\ref{sec1}, we present the quantum Otto cycle and the relevant thermodynamics quantities necessary to characterize the Otto heat engine and refrigerator. In Sec.~\ref{sec:II}, we calculate these thermodynamics quantities for the KNO Otto machine. Results are presented in Sec.~\ref{KNOheat} for the KNO operating as Otto heat engine and in Sec.~\ref{RefriOtto} operating as an Otto refrigerator. Final remarks are presented in Sec.~\ref{sec:V}.

\section{Quantum Otto Cycle \label{sec1}}

\begin{figure}[t]
\begin{center}
\includegraphics[width=0.4\textwidth]{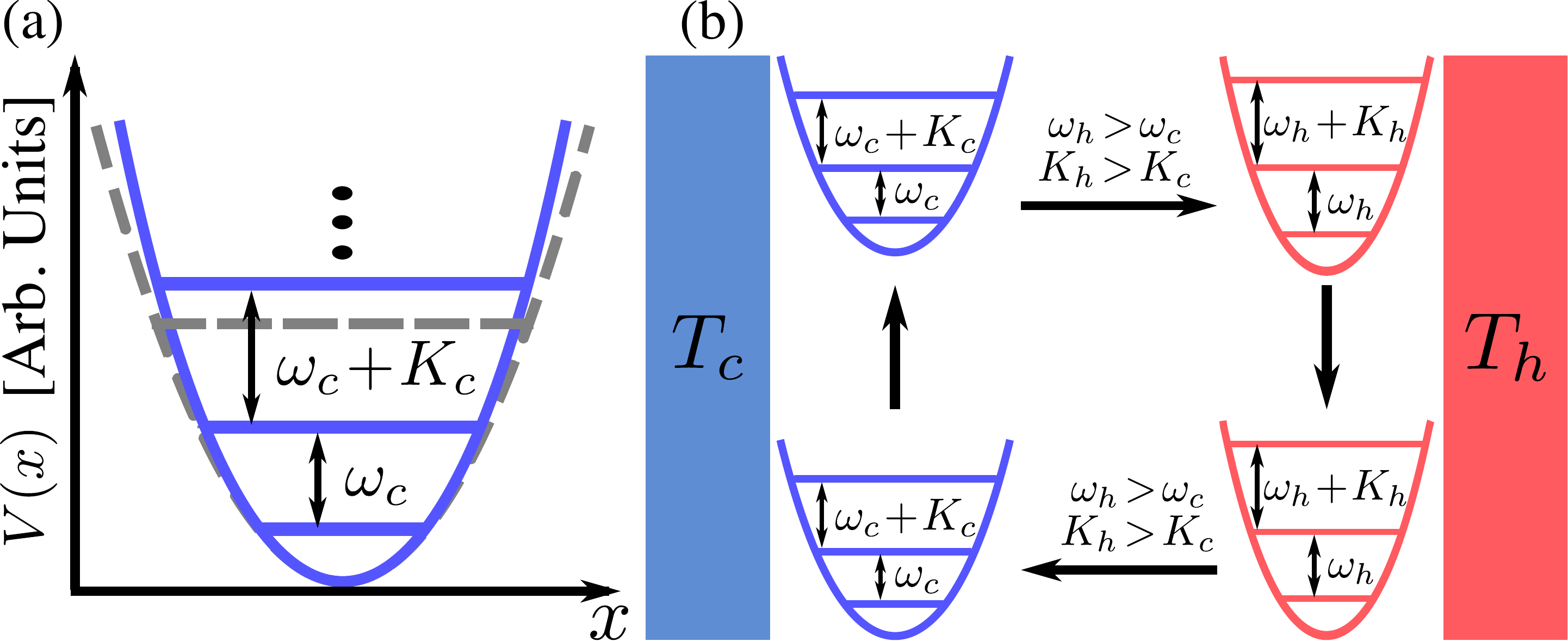}
\caption{ (a) Comparison of the quantum harmonic potential (dashed line) and the Kerr-nonlinear oscillator (solid lines) potential and energy levels. Kerr-nonlinearity makes energy levels to be unevenly spaced and dependent on the boson number excitation $n$. This is in stark contrast with quantum harmonic oscillator, which presents an equally spaced energy level structure. (b) The Kerr-nonlinear oscillator operating in the Otto cycle. During the cooling and heating strokes, the KNO is, respectively, in contact with the cold- and hot-thermal reservoirs. In the expansion and compression strokes the KNO is disconnected from the thermal reservoirs and evolves unitarily while varying either the frequency $\omega_h \leftrightarrow \omega_h$ or Kerr-nonlinear strength $K_c \leftrightarrow K_h$.  \label{fig1}}
\end{center}
\end{figure}

Our model consists of a KNO as the working substance and interacting with two thermal reservoirs at different temperatures to implement a quantum Otto machine (QOM). We show by varying the KNO frequency and the nonlinear coupling strength that the QOM can be controlled to either extract or perform work. The QOM is formed by two isochoric and two isentropic branches. In the first isochoric branch, the KNO is coupled to a cold-thermal reservoir, while in second branch, it is coupled to a hot-thermal reservoir. In the isentropic branches, the KNO is decoupled from the thermal reservoirs and left to evolve unitarily to complete the cycle \cite{Kosloff17}. The four strokes forming the QOM are illustrated in Fig.~\ref{fig1}(b) and described in detail below

\emph{ i) Cooling stroke:} The KNO is weakly coupled to a cold thermal reservoir up to thermalization. The KNO-thermalized state is described by the Gibbs state $\rho_{1} = e^{-\beta_{c} H_{c}}/\text{Tr}\left(\text{e}^{-\beta_{c} H_{c}}\right)$,
with $H_{c}$ being the KNO Hamiltonian and $\beta_{c}=1/k_{B}T_{c}$, with $k_{B}$ being the Boltzmann constant and $T_{c}$ the temperature of the cold reservoir.

\emph{ ii) Expansion stroke.} In this step, the KNO evolves unitarily from the state $\rho_{1}$ to $\rho_{2}=U(t)\rho_{1}U^{\dagger}(t)$, with $U(t)$ the unitary time-evolution operator. During the time evolution, the KNO Hamiltonian evolves from $H_{c}$ to $H_{h}$. As we are interested in obtaining the ultimate limit achievable by the Otto cycle, we will consider a quasi-static process. In the quasi-static limit, the specific form of the unitary time-evolution operator $U(t)$ is not important and its effect is to change the KNO frequency and nonlinearity strength.

\emph{ iii) Heating stroke.} In this branch, the KNO is weakly coupled to a hot-thermal reservoir at temperature $T_{h}$ until reaching the thermal Gibbs state $\rho_{3}=\text{e}^{-\beta_{h}H_{h}}/\text{Tr}\left(\text{e}^{-\beta_{h}H_{h}}\right)$,
with $\beta_{h}=1/k_{B}T_{h}$.

\emph{ iv) Compression stroke.} This last step is accomplished by reversing the protocol employed to perform the expansion stroke, such that the KNO Hamiltonian transforms from $H_{h}$ to $H_{c}$ and its state evolves unitarily from $\rho_{3}$ to $\rho_{4}=U^{\dagger}\rho_{3}^{S}U(t)$.

In the complete cycle, work is either produced or consumed in the the expansion and compression strokes. Thus, the net work is defined as $W=W_{1\rightarrow2}+W_{3\rightarrow4}$, with $W_{1\rightarrow2}$ the work in the expansion stroke and $W_{3\rightarrow 4}$ the work in the compression stroke. In these strokes, the QOM does not exchange heat with the thermal reservoirs. Thus, from the first law of thermodynamics, the net work is given by the variation of total energy \cite{callen}. Following Ref.~[\onlinecite{key-26Alicki79}], the net work is defined as 
\begin{equation} \label{eq:W}
W= \tr \left(\rho_{2}H_{h}\right)-\tr \left(\rho_{1}H_{c}\right)+\tr \left(\rho_{4}H_{c}\right)- \tr\left(\rho_{3}H_{h}\right). 
\end{equation}

In the cooling and heating strokes, the QTM does not perform or extract work. However, it exchanges heat with the thermal reservoirs. In the cooling stroke, the KNO is coupled to the cold-thermal reservoir and the heat exchanged $Q_{c} = Q_{4\rightarrow1}$ takes the form
\begin{equation} \label{Qc}
Q_{c} = \tr\left(\rho_{1}H_{c}\right)-\tr\left(\rho_{4}H_{c}\right).
\end{equation}
In the heating stroke, the heat exchanged $Q_{h}=Q_{2\rightarrow3}$ between the KNO and the hot-thermal reservoir is 
\begin{equation} \label{Qh}
Q_{h}=\tr \left(\rho_{3}H_{h}\right)-\tr \left(\rho_{2}H_{h}\right).
\end{equation}

In the next section, the above formulas will be used to calculate the net work and heat for the KNO Otto cycle. Then, for a given set of parameters, we demonstrate that the KNO thermal machine can behave either as a quantum engine or as a refrigerator.

\section{\label{sec:II} Work and Heat in the Otto Cycle of a Kerr-Nonlinear Oscillator}

We now calculate the net work $W$ and the heats, $Q_c$ and $Q_h$, corresponding to the Otto cycle [see Fig. \ref{fig1}(b)] described in the previous section. The working substance is formed by a quantum Kerr-nonlinear oscillator. The quantum KNO Hamiltonian responsible for extracting work from the Otto cycle is
\begin{equation}
H_{c}= \hbar \omega_{c} a^{\dagger}a +\hbar \frac{K_c}{2} a^{\dagger2}a^{2}.
\end{equation}
The first term of the Hamiltonian describes a single-mode harmonic oscillator of frequency $\omega_{c}$, with $a^{\dagger}$ and $a$ the creation and annihilation operators, respectively. The last term is the non-linear interaction of strength $K_c$. Here, $\hbar$ is the reduced Planck constant $h/2\pi$. Kerr Hamiltonians are easily engineered in circuit QED \cite{blais-rmp2021,brock-prapp2021} and cavity optomechanics \cite{aldana-pra2013} devices. These quantum devices are operated in the microwave regime. To produce work, either the frequency or the nonlinear strength of the KNO medium should vary. In circuit QED devices, the frequency and Kerr strength can be tuned by varying the capacitor charge or flux passing through a SQUID loop forming the resonator \cite{brock-prapp2021}, thus, modifying the above Hamiltonian to
\begin{equation}
H_{h}=\hbar \omega_{h} a^{\dagger}a +\hbar \frac{K_{h}}{2} a^{\dagger2}a^{2}.
\end{equation}

We are now in the position to compute the net work [Eq.~\eqref{eq:W}] and the heats [Eqs.~\eqref{Qc} and \eqref{Qh}] in the Otto cycle described in the Sec. \ref{sec1}. We start by computing $\tr\left(\rho_{2}H_{h}\right)$, with $\rho_{2}=U(t)\rho_{1}U^{\dagger}(t)$ and $\rho_{1} = e^{-\beta_{c} H_{c}}/\text{Tr}\left(e^{-\beta_{c} H_{c}} \right)$:
\begin{equation} \label{eq:6}
\tr(\rho_{2}H_{h}) =\hbar \tr \sum_{n=0}^{\infty} p_{n}^{c} U(t) |n \rangle \langle n | U^\dagger (t) \left[ \omega_{h} a^{\dagger}a +\frac{K_{h}}{2} a^{\dagger 2}a^{2} \right], 
\end{equation}
with $p_{n}^{c}=e^{-\hbar \beta_{c}\left[\omega_{c}n+K_{c}(n^{2}-n)/2 \right] }/Z_{T_{c}}$ the population of the $n$-Fock state, and the partition function $Z_{T_{c}}=\sum_{n=0}^{\infty}\exp\left\{-\hbar \beta_{c}\left[\omega_{c} n + K_{c}(n^{2}-n)/2\right]\right\}$. Using bosonic commutation relations and inserting the completeness relation $\sum_m |m\rangle \langle m|  = 1$, Eq.~\eqref{eq:6} takes the form
\begin{equation} \label{eq:rho2}
\tr\left(\rho_{2}H_{h}\right)= \hbar \sum_{n,m=0}^{\infty} p_{n}^{c} U_{n,m}(t) \left[\omega_{h}m +\frac{K_{h}}{2}(m^{2}-m)\right],
\end{equation}
where we defined $U_{n,m}(t) =  \left|\langle n \right|U(t) \left|m \rangle \right|^{2}$. Analogously, the traces in Eqs.~\eqref{eq:W}, \eqref{Qc} and \eqref{Qh} are equal to 

\begin{subequations}
\begin{align}  \label{eq:rho1}
\tr\left(\rho_{1}H_{c}\right) &=\hbar \sum_{n=0}^{\infty}p_{n}^{c}\left[\omega_{c}n+K_c (n^{2}-n)/2\right],
\end{align}
\begin{align}  \label{eq:rho3}
\tr\left(\rho_{3}H_{h}\right)&=\hbar \sum_{n,m=0}^{\infty}p_{n}^{h} U_{n,m}(t) \left[\omega_{h}m+ \frac{K_{h}}{2}(m^{2}-m)\right],
\end{align}
\begin{align}  \label{eq:rho4}
\tr\left(\rho_{4}H_{c}\right) &=\hbar \sum_{n=0}^{\infty}p_{n}^{h}\left[\omega_{c}n+K_c(n^{2}-n)/2\right],
\end{align}
\end{subequations}
where we defined $p_{n}^{h}=e^{ -\hbar \beta_{h}\left[\omega_{h}n+ K_{h}(n^{2}-n )/2\right] }/Z_{T_{h}}$, and the partition function $Z_{T_{h}}=\sum_{n=0}^{\infty}\exp \left\{-\hbar \beta_{h}\left[\omega_{h}n+K_{h}(n^{2}-n )/2 \right]\right\}$.

To calculate the net work $W$ and heats $Q_c$ and $Q_h$, we need to specify the time-evolution operator $U(t)$. As we are interested in the ultimate limit achievable by the Otto cycle, we consider a quasi-static process, for which the system is always in an eigenstate of the instantaneous Hamiltonian. In this limit, the form of $U(t)$ is not important and, according to the adiabatic theorem, $\left\langle n\right|U(t)\left|m\right\rangle =\delta_{n,m}$. We replace this result in Eqs.~\eqref{eq:rho1},~\eqref{eq:rho3} and~\eqref{eq:rho4} to compute the net work defined in Eq.~\eqref{eq:W}
\begin{equation}
W=-\hbar \sum_{n=0}^{\infty} \Delta p_n \left[\Delta \omega n +\frac{\Delta K}{2} \left(n^{2}-n\right) \right],\label{WW}
\end{equation}
with $\Delta p_n = p_{n}^{h} - p_{n}^{c}$ the population difference, $\Delta \omega = \omega_{h} - \omega_{c}$, and $\Delta K =K_{h} - K_{c}$. Similarly, we obtain the heat exchanged between the KNO and the reservoirs during Otto cycle. 
Replacing Eqs.~\eqref{eq:rho1} and \eqref{eq:rho4} into Eq.~\eqref{Qc} we obtain
\begin{equation}
Q_{c}=-\hbar \sum_{n=0}^{\infty} \Delta p_n \left[\omega_{c} n + K_{c}(n^{2}-n)/2\right].\label{Qc1}
\end{equation}
Substituting Eqs.~\eqref{eq:rho2} and \eqref{eq:rho3} in Eq.~\eqref{Qh} gives
\begin{equation}
Q_{h}=\hbar \sum_{n=0}^{\infty}\Delta p_n \left[\omega_{h} n + K_{h}(n^{2}-n )/2 \right].\label{Qh1}
\end{equation}

From Eqs.~\eqref{WW}, \eqref{Qh1} and (\ref{Qc1}) we observe that all three quantities depend on the population difference $\Delta p_n$. Thus, a thermal population imbalance will generate work and heat. Besides, net work is produced or extracted by the variation of the Hamiltonian parameters, as expected.

\section{Results \label{secRes}}

In this section, we will use the above definitions of work and heats to calculate the heat engine efficiency and the refrigerator performance of the KNO Otto engine. We demonstrate that KNO heat engine efficiency and refrigerator performance outperforms its counterpart formed by a simple QHO.

\subsection{Heat Engine Powered by Kerr Nonlinearity \label{KNOheat}}

To build a quantum Otto engine based on the Otto cycle described in Sec.~\ref{sec1}, we first need to assure that the engine conditions are fulfilled. In the case of QOM, the engine exists when work is performed by the KNO while absorbing heat from the hot reservoir. Therefore, we search for a parameter regime in which $W<0$ [Eq.~\eqref{WW}], $Q_{h}>0$ [Eq.~\eqref{Qh1}] and  $Q_{c}<0$ [Eq.~\eqref{Qc1}] are satisfied simultaneously. In this regime, a KNO-based Otto heat engine exist and will be compared to both QHO-based Otto and Carnot machines.
\begin{figure}[!t]
\begin{center}
\includegraphics[width=0.45\textwidth]{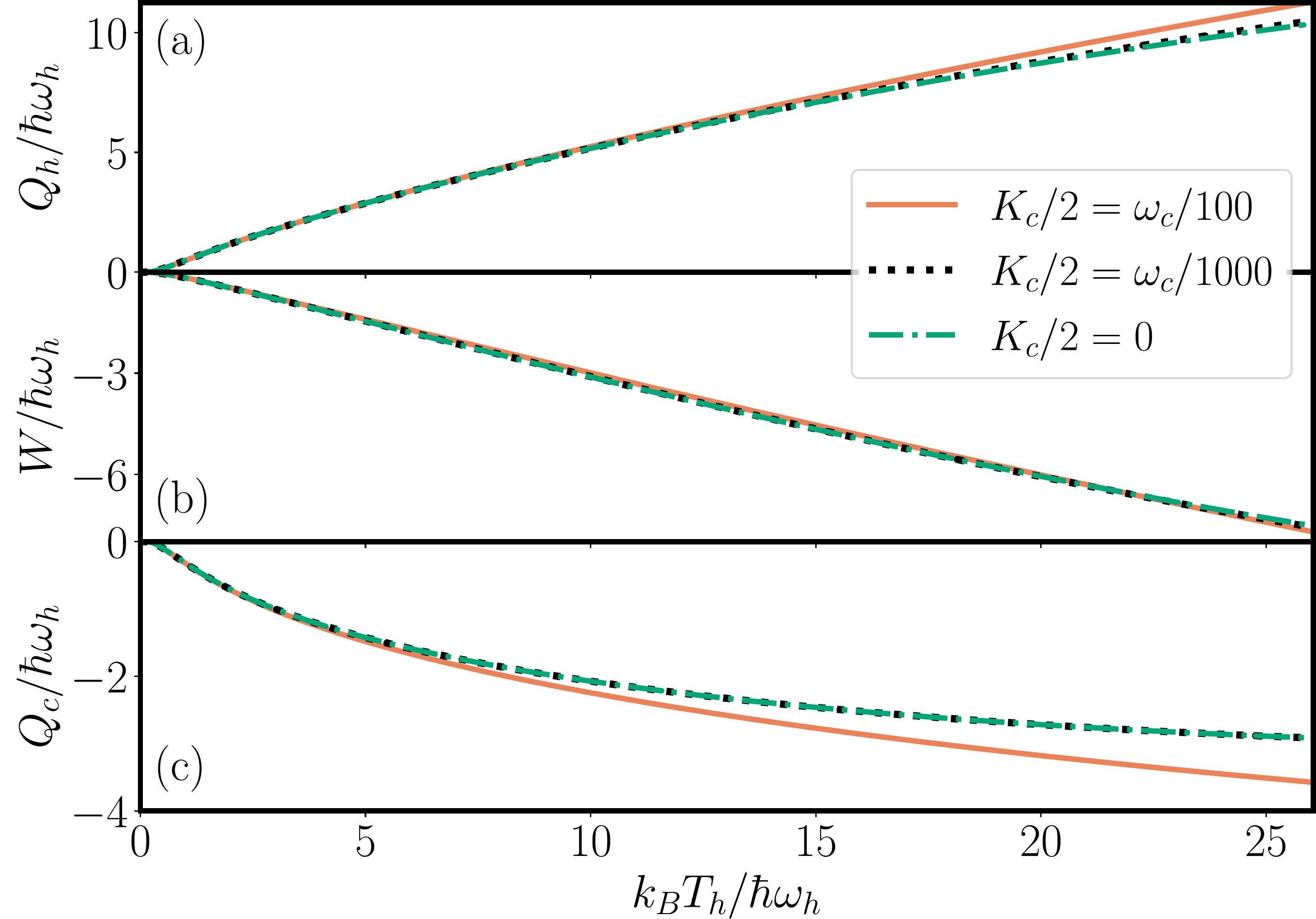}
\caption{(a) heat $Q_h$, (b) net work $W$, and (c) heat $Q_c$ as a function of the hot-reservoir temperature for values of the Kerr-nonlinearity strength $K_c/2 = \omega_c/1000$ (solid line) , $K_c/2 = \omega_c/100$ (dotted line) and $K_c = 0$ (dash dot line). The other parameters are fixed and equal to $\omega_h = 2\pi \times 4$~GHz,  $\omega_c = 0.7 \omega_h$, $T_c = 0.1 T_h$ and $K_h/2 = 0.1 \omega_h$. The heat engine conditions, $W<0$, $Q_{h}>0$ and $Q_{c}<0$, are satisfied for wide range of temperatures. \label{fig2}}
\end{center}
\end{figure}

Figure~\ref{fig2} illustrates the (a) heat $Q_h$, (b) net work $W$, and (c) heat $Q_c$ as a function of the hot-reservoir temperature for different values of the Kerr-nonlinearity strength $K_c/2 = \omega_c/100$ (solid line), $K_c/2 = \omega_c/1000$ (dotted line) and $K_c = 0$ (dash dot line). We consider $\omega_h = 2\pi \times 4$~GHz,  $\omega_c = 0.7 \omega_h$, $T_c = 0.1 T_h$ and $K_h/2 = 0.1 \omega_h$ as fixed parameters. For these parameters, the engine condition is satisfied and we observe that the net work $W$, $Q_h$ and $Q_c$ are weakly dependent on the Kerr-nonlinearity strength $K_c$. Moreover, the net work $|W|$ is always smaller than $Q_h$, indicating that the quantum Otto engine efficiency is smaller than 1, as expected.

We are now in position to compute the heat engine efficiency $\eta=- W/Q_{h} $ for our KNO based Otto machine. Using Eqs.~\eqref{eq:W} and \eqref{Qh} the efficiency is written as
\begin{equation} \label{Effic}
\eta=1-\frac{\omega_{c}}{\omega_{h}}\left\{ \frac{\sum_{n=0}^{\infty}\Delta p_n \left[ n + \frac{K_{c}}{2\omega_{c}}\left(n^{2}-n\right)\right]}{\sum_{n=0}^{\infty}\Delta p_n \left[ n + \frac{K_{h}}{2\omega_{h}}\left(n^{2}-n\right)\right]}\right\}.
\end{equation}
The above efficiency reduces to the QHO Otto efficiency $\eta_{HO}=1- \omega_{c}/\omega_{h}$ when $K_{c}/\omega_{c}= K_{h}/\omega_{h}$ or, as expected, in the absence of Kerr non-linearity $K_{c}=K_{h}=0$. The best strategy to maximize the efficiency is by minimizing the numerator while maximizing the denominator in Eq.~\eqref{Effic}.  This maximization is accomplished, for example, by taking $K_{c}=0$ and letting $K_{h}\rightarrow\infty$. However,  $K_{c}$ is constrained by experimental feasibility. For instance, in the context of circuit QED, $K_{h}/\omega_h$ can vary from 0.001 to 0.1 \cite{blais-rmp2021}, and in a novel superconducting device termed \textit{quarton} the non-linearity strength can be made as large as $1/3$ of its natural frequency \cite{yan-arxiv2020}. Moreover, the non-linear strength $K_h$ can be externally tuned through gate voltage or flux bias \cite{brock-prapp2021}, thus, allowing to maximize the heat engine efficiency.
\begin{figure}[!t]
\begin{center}
\includegraphics[width=0.45\textwidth]{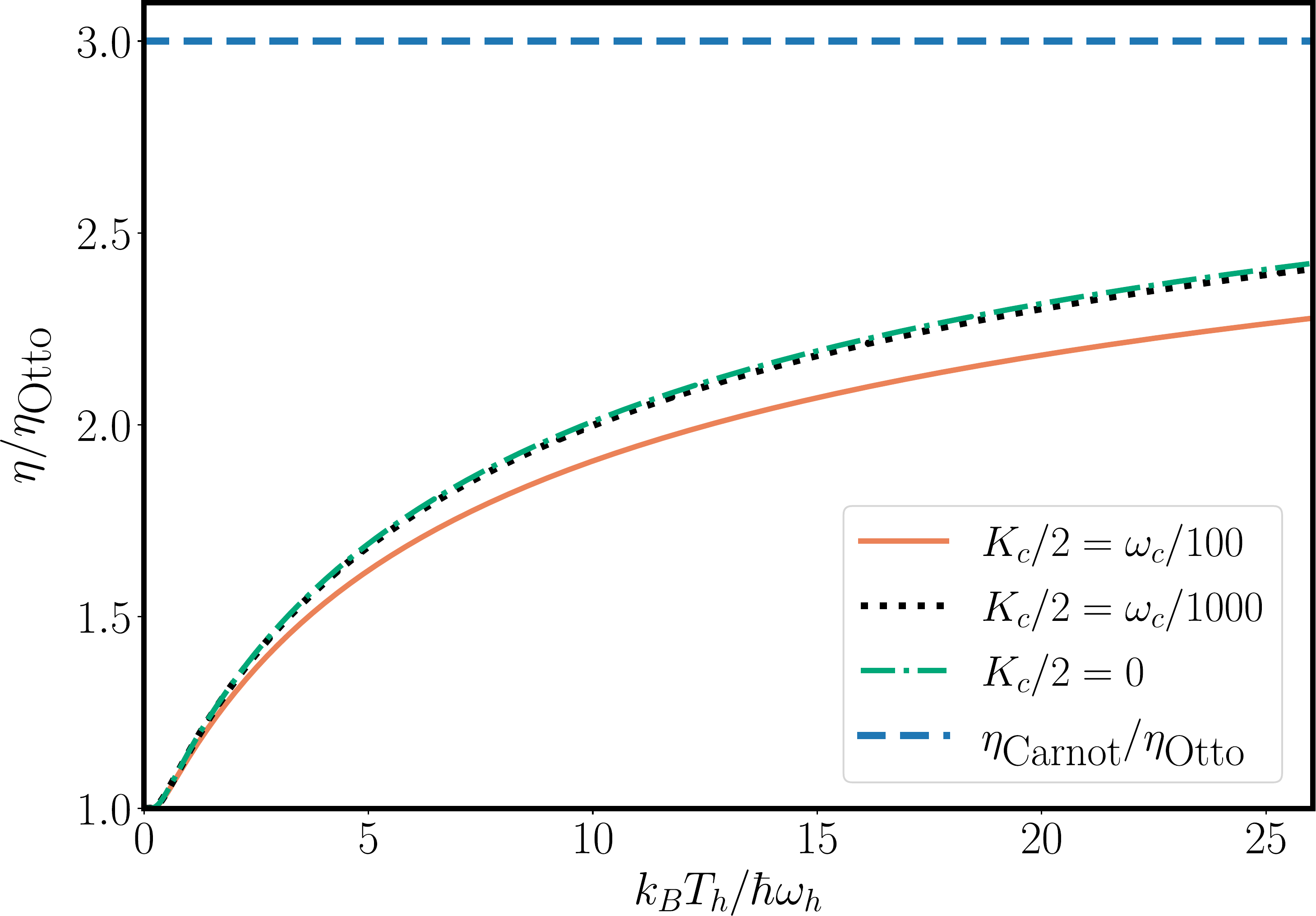}
\caption{ Heat engine efficiency $\eta/\eta_\text{Otto}$ as a function of the hot-reservoir temperature for values of the Kerr-nonlinearity strength $K_c/2 = \omega_c/1000$ (solid line), $K_c/2 = \omega_c/100$ (dotted line) and $K_c = 0$ (dash dot line). The parameters $\omega_h = 2\pi \times 4$~GHz,  $\omega_c = 0.7 \omega_h$, $T_c = 0.1 T_h$ and $K_h/2 = 0.1 \omega_h$ are fixed. The KNO heat engine surpasses the quantum harmonic oscillator efficiency for a wide range of temperatures $T_h$ and it is weakly dependent on the value of cold-reservoir Kerr strength $K_c$. The efficiency $\eta/\eta_\text{Otto}$ is limited by the Carnot's efficiency $\eta_\text{Carnot}/\eta_\text{Otto} = [1-T_{c}/T_{h}]/[1-\omega_c/\omega_h]=3 $ (dashed line), but for this set of parameters it always surpasses the QHO Otto efficiency $\eta_\text{Otto} = 1-\omega_c/\omega_h = 0.3$. \label{fig3}}
\end{center}
\end{figure}

Figure~\ref{fig3} shows the KNO heat engine efficiency $\eta$ as a function of the hot-reservoir temperature for the same set of parameters used in Fig.~\ref{fig2}. We observe that the efficiency increases with increasing temperature $T_h$, reaching an efficiency of approximately 75\% for high temperatures. Furthermore, the efficiency $\eta/\eta_\text{Otto}$ of our quantum Otto KNO is always larger than one, thus, indicating the the KNO efficiency surpasses the QHO Otto efficiency $\eta_\text{Otto} = 1-\omega_c/\omega_h$. However, the KNO efficiency $\eta/\eta_\text{Otto}$ does not surpass the Carnot efficiency $\eta_\text{Carnot}/\eta_{Otto} =[1-T_{c}/T_{h}]/[1-\omega_c/\omega_h]=3$ (dashed line). We also note that $\eta$ depends weakly on the cold-reservoir Kerr non-linearity strength. Indeed, for $K_c < K_h$, the efficiency is always larger than $\eta_\text{Otto}$. For $K_c > K_h$, the heat engine condition $W<0$ and $Q_{h}>0$ is only satisfied at low temperatures, but the heat engine efficiency is smaller than $\eta_\text{Otto}$ [not shown in Fig.~\ref{fig3}].

We observe in Fig.~\ref{fig3} that KNO Otto heat engine efficiency can reach approximately 2.5 times the QHO heat engine's efficiency. Thus, these results show that the Kerr non-linear interaction enhances the efficiency of heat engines in comparison to QHO heat engines. It is important to emphasize that, in the context of circuit QED, both the frequency and Kerr non-linearity strength can be tuned simultaneously \cite{brock-prapp2021}.

%


\subsection{\label{sec:III}Refrigerator Powered by Kerr Nonlinearity \label{RefriOtto}}

\begin{figure}[t]
\begin{center}
\includegraphics[width=0.45\textwidth]{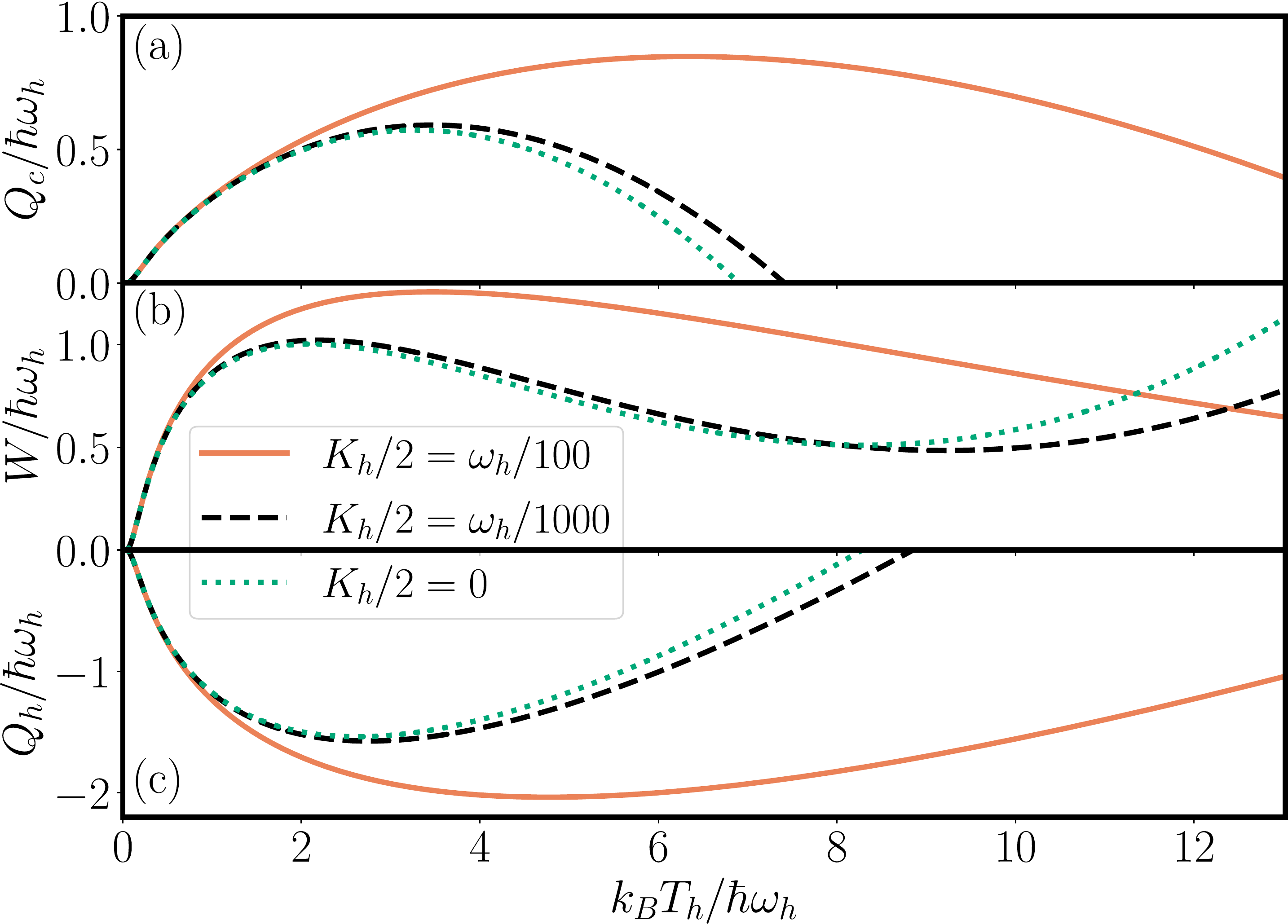}
\caption{(a) heat $Q_c$, (b) net work $ W$ and (c) heat $Q_h$ as a function of the hot-reservoir temperature $T_h$ for different values of the Kerr nonlinearity strength $K_h/2 = \omega_h/100$ (solid line), $K_h/2 = \omega_h/1000$ (dash line) and $K_h/2 = 0$ (dotted line). The parameters $\omega_h=  2\pi \times  8$ GHz, $\omega_c = 2\pi \times 1.6$ GHz, $K_c/2 = \omega_c/10$ and $T_c = 0.7 T_h$ are fixed. The refrigerator condition $W>0$, $Q_c>0$ and $Q_h<0$ is satisfied for for low-temperatures $T_h$. \label{fig4}}
\end{center}
\end{figure}

In this section, we extend the formalism presented in the previous sections to investigate the role of Kerr nonlinearity in the coefficient of performance $\epsilon$ of a quantum Otto refrigerator. Refrigerators operate by consuming work to extract heat from the cold reservoir and pump it into the hot reservoir \cite{callen}. Thus, refrigeration is achieved in a quantum Otto cycle when $W>0$ [Eq.~\eqref{WW}], $Q_c>0$ [Eq.~\eqref{Qc1}] and $Q_h<0$ [Eq.~\eqref{Qh1}] for a fixed set of parameters. Fig.~\ref{fig4} illustrates (a) $Q_c$, (b) $ W$ and (c)  $Q_h$ as a function of the hot-reservoir temperature $T_h$ for different values of the Kerr nonlinearity strength $K_h/2 = \omega_h/100$ (solid line), $K_h/2 = \omega_h/1000$ (dash line) and $K_h/2 = 0$ (dotted line). To obtain the results in Fig.~\ref{fig4}, we considered $\omega_h=  2\pi \times  8$ GHz, $\omega_c = 2\pi \times 1.6$ GHz, $K_c/2 = \omega_c/10$ and $T_c = 0.7 T_h$. Fig.~\ref{fig4} shows that the net work $W$ and heat $Q_c$ ($Q_h$) have a concave (convex) behavior with increasing $T_h$, and the refrigerator condition is only satisfied at low temperatures $T_h$. This behavior differs from the KNO Otto heat engine [Fig.~\ref{fig3}], which shows increase efficiency with increasing temperature $T_h$ [see Fig.~\ref{fig3}].

To further characterize the KNO Otto refrigerator, we calculate the coefficient of performance $\epsilon= Q_C /W $, which is defined as the ratio of heat transferred from the cold reservoir $Q_C$ to the net work $W$ performed in the cycle  \cite{abah-epl-2016}. Using Eqs.~\eqref{WW} and \eqref{Qc1}, the coefficient of performance takes the form
\begin{equation} \label{COP}
\epsilon =\frac{\omega_{c}}{\triangle\omega}\frac{\sum_{n=0}^{\infty}\Delta p_n\left[ n+\frac{K_{c}}{2\omega_{c}}\left(n^{2}-n\right)\right]}{\sum_{n=0}^{\infty}\Delta p_n \left[n +\frac{\Delta K}{2 \Delta \omega}\left(n^{2}-n\right)\right]}.
\end{equation}
The QHO Otto coefficient of performance $\epsilon_\text{Otto} = \omega_{c}/\Delta \omega$ is recovered for $K_{c}/\Delta K = \omega_{c}/\Delta \omega $ and, as expected, in the absence of Kerr non-linearities $K_{c}=K_{h}=0$. In order to maximize the $\epsilon$, the ratio $K_{c}/\omega_{c}$ must be made as large as possible, while taking the limit of $\Delta K / \Delta \omega$ going to zero.
\begin{figure}[t]
\begin{center}
\includegraphics[width=0.45\textwidth]{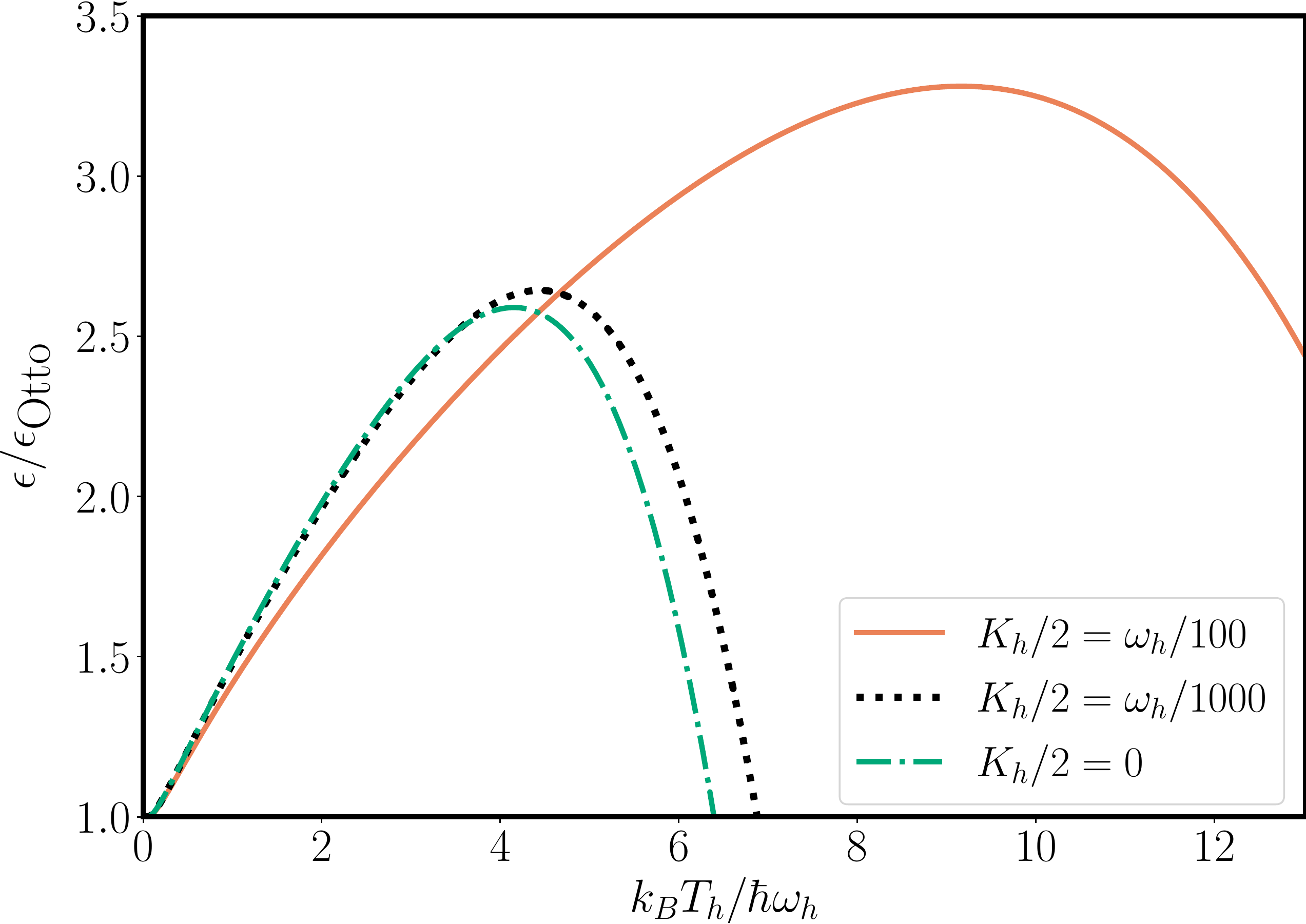}
\caption{Coefficient of performance $\epsilon/\epsilon_\text{Otto}$ as a function of the hot-reservoir temperature $T_h$ for different values of Kerr non-linearity strength $K_h/2 = \omega_h/100$ (solid line), $K_h/2 = \omega_h/1000$ (dash line) and $K_h/2 = 0$ (dotted line). We consider the parameters $\omega_h=  2\pi \times  8$ GHz, $\omega_c = 2\pi \times 1.6$ GHz, $K_c/2 = \omega_c/10$ and $T_c = 0.7 T_h$ to be fixed. At finite temperature, $\epsilon$ is always lager than the coefficient of performance of QHO refrigerator $\epsilon_\text{Otto} = \omega_{c}/\Delta \omega = 1/3$ and smaller than coefficient of performance of the Carnot's refrigerator $\epsilon_\text{Carnot} = T_c/(T_h - Tc) = 7/3$. \label{fig5}}
\end{center}
\end{figure}

Fig~\ref{fig5} shows the coefficient of performance $\epsilon/\epsilon_\text{Otto}$ as a function of the hot-reservoir temperature for the same set of parameters used to obtain Fig.~\ref{fig4}. The coefficient of performance $\epsilon$ first increases with increasing hot-reservoir temperature and achieves a maximum at temperature $T_h^*$. Contrary to KNO heat engine efficiency, the KNO refrigerator performance only surpasses the QHO refrigerator $\epsilon_\text{Otto} = \omega_{c}/\Delta \omega = 1/3$  at low temperatures. Similar to the net work and heat $Q_C$, the coefficient of performance present a concave dependence with increasing $T_h$. Moreover, similarly to heat engines, $\epsilon$ is always smaller than $\epsilon_\text{Carnot} = T_c/(T_h - Tc) = 7/3$.

We observe in Fig.~\ref{fig5} that, for certain parameters, the improvement in performance can exceed 3 times the ideal performance $\eta_\text{Otto}$ of the quantum HO refrigerator. Thus, the above results demonstrate that the KNO Otto refrigerator can be implemented in a circuit QED device and its performance easily surpasses the performance of the quantum HO refrigerator in a Otto cycle.

%

\section{\label{sec:V} Final Remarks}

We investigated the role played by Kerr nonlinearity powering quantum thermal machines operating in Otto cycle. By using realistic parameters taken from circuit QED \cite{blais-rmp2021,yan-arxiv2020,brock-prapp2021}, we demonstrated that both the engine efficiency $\eta$ and the refrigerator coefficient of performance $\epsilon$ are enhanced, surpassing their quantum harmonic oscillator counterparts $\eta_\text{Otto}$ and $\epsilon_\text{Otto}$ and limited by the Carnot's efficiency and coefficient of performance. Indeed, for the parameters used, taking into account the experimental feasibility in the context of the circuit QED, Kerr nonlinearity enables to achieve gains of up to 2.5 times for heat engines and above 3 times for refrigerators in comparison with the QHO Otto machines. Differently from the quantum harmonic oscillator Otto machines, KNO Otto thermal machines present a temperature dependence. Finally, our study shows that nonlinear effects are important and can lead to new interesting features in Otto machines. For instance, the nonlinear interaction in a Josephson parametric amplifier \cite{blais-rmp2021}, which is responsible for parametric amplification and squeezing is been currently investigated \cite{mattos2021}.


\section*{Acknowledgments }

We acknowledge financial support from the Brazilian agencies CAPES (Financial code 001), CNPq and FAPEG. This work was performed as part of the Brazilian National Institute of Science and Technology (INCT) for Quantum Information (Grant No. 465469/2014-0). U.C.M. acknowledges the support from CNPq-Brazil (Project No. 309171/2019-9).


%

\end{document}